\documentclass{article}

\usepackage{amsfonts}
\usepackage{dsfont}
\usepackage{amssymb}
\usepackage{amsmath}
\usepackage[bb=boondox]{mathalfa}
\usepackage{amsxtra}
\usepackage{graphicx}
\usepackage{psfrag}
\usepackage{mathrsfs}
\usepackage{multirow}
\usepackage{stmaryrd}
\usepackage{subfigure}
\usepackage{empheq}
\usepackage[normalem]{ulem}
\usepackage{pdfpages}
\usepackage{wrapfig}
\usepackage{longtable}
\usepackage{footmisc}
\usepackage{hyperref}
\usepackage{lineno}
\usepackage{outlines}
\usepackage{enumitem}
\usepackage{lipsum}
\usepackage{upgreek}
\usepackage{bm}
\usepackage{scalerel}

\usepackage{float}
\usepackage{caption} 
\usepackage[margin=1.5cm]{geometry}
\usepackage{soul}
\usepackage{dblfloatfix}
\usepackage{siunitx}
\usepackage{tikz}
\usepackage{sectsty}
\usepackage{booktabs}
\usepackage{algorithm}
\usepackage{algpseudocode}
\usepackage{breqn}
\usepackage{bbold}
\usepackage[backend=biber,sorting=none]{biblatex}
\subsectionfont{\fontsize{14}{15}\selectfont}

\usepackage[nameinlink]{cleveref}
\crefname{subsection}{subsection}{subsections}
\crefname{appendix}{}{} 

\newcommand{\ignore}[1]{}

\newcommand{\up}[1]{\mathbf{#1}}

\DeclareMathOperator*{\argmin}{\arg\!\min}

\algnewcommand\And{\textbf{and}}

\newcolumntype{L}{>{$}l<{$}}

\newcommand{\dd}{{\rm d}}
\newcommand{\fracd}[2]{\frac{\dd #1}{\dd #2}}

\newcommand{\transpose}{\intercal} 
\newcommand{\sym}[1]{{\rm{symm}} \left[ #1 \right] }

\newcommand{\quant}{{\up{A}}}
\newcommand{\pkone}{\bm{\Uppi}}
\newcommand{\pktwo}{{\up{S}}}
\newcommand{\cauchy}{\bm{\upsigma}}

\newcommand{\rotkir}{{\up{T}}}
\newcommand{\mathencky}{{\up{H}}}
\newcommand{\spathencky}{\bm{\upeta}}

\newcommand{\devH}{\mathencky_0}

\newcommand{\barI}{\overline{I}}
\newcommand{\varu}{\delta{\bm{u}}}
\newcommand{\grad}{\boldsymbol \nabla}

\newcommand{\trace}[1]{ {\rm tr}\left[ #1 \right] }

\newcommand{\stiffA}{\mathbb{A}} 
\newcommand{\IIsym}{{\mathbb{I}^{\, \rm sym}}}

\newcommand{\deformmap}{\bm{\varphi}}

\providecommand{\keywords}[1]
{
  \small	
  \textit{Keywords}: #1
}

\def\bnabla{\mbox{\boldmath$\nabla$}}

\def\bP{\mbox{\boldmath$ P$}}

\def\bX{\mbox{\boldmath$ X$}}

\def\bt{\mbox{\boldmath$ t$}}
\def\bu{\mbox{\boldmath$ u$}}

\def\bx{\mbox{\boldmath$ x$}}

\def\shrug{\texttt{\raisebox{0.75em}{\char`\_}\char`\\\char`\_\kern-0.5ex(\kern-0.25ex\raisebox{0.25ex}{\rotatebox{45}{\raisebox{-.75ex}"\kern-1.5ex\rotatebox{-90})}}\kern-0.5ex)\kern-0.5ex\char`\_/\raisebox{0.75em}{\char`\_}}}
\numberwithin{equation}{section}

\title{Variation-matching sensitivity-based virtual fields for hyperelastic material model calibration}

\author{Denislav~P.~Nikolov$^{1}$, Zhiren~Zhu$^{1}$, Jonathan~B.~Estrada$^{1,*}$ \\
\small $^{1}$ Department of Mechanical Engineering, University of Michigan, 2350 Hayward St., Ann Arbor, 48109, MI, United States \\
\small $^{*}$ Corresponding author \\
\small Email address: jbestrad@umich.edu}

\date{}

\begin{document}
\maketitle

\begin{abstract}
\noindent
Accurate identification of nonlinear material parameters from three-dimensional full-field deformation data remains a challenge in experimental mechanics.
The virtual fields method (VFM) provides a powerful, computationally efficient approach for material model calibration, however, its success depends critically on the choice of virtual fields and the informativeness of available kinematic data.
In this work, we advance the state-of-the-art discrete formulation of the sensitivity-based virtual fields (SBVF) method by systematically developing and comparing alternative variational and analytical SBVFs within a strain-invariant-based modeling framework.

\medskip
\noindent A central contribution of this work is the implementation and assessment of variation-based SBVFs (vSBVFs), formulated using directional G\^ateaux derivatives, as well as virtual fields derived from analytical differentiation (aSBVFs) which provide explicit, model-tailored virtual displacement fields for parameter identification.
Using simulated noisy volumetric datasets, we demonstrate that vSBVFs and aSBVFs enable procedural, automated construction of optimal virtual fields for each material parameter, substantially enhancing the robustness and efficiency of calibration without the need for manual field selection or high temporal resolution in the data acquisition.
We quantify data richness---the effective diversity of sampled kinematic states---showing that increased data richness via sample geometry and loading protocols leads to improved parameter identifiability.
These findings establish a pathway for automated, noise-robust material model calibration suitable for future deployment with experimental full-field imaging of soft, complex materials, and provide a foundation for optimizing shape topology and extending to viscoelastic and anisotropic behaviors.
\end{abstract}


\section{Introduction}
\label{sec:intro}

Full-field measurement techniques like digital image correlation (DIC) \cite{sutton_DICBook_2009} and displacement-encoded magnetic resonance cartography (MR-\textbf{\textit{u}}) \cite{aletras_DENSE_1999,neu_dense_2008,scheven_physicsMRu_2020} have been used to acquire full-field kinematics of increasingly complex boundary value problems (BVP) \cite{estrada_MRu_2020,luetkemeyer_bioMRu_2021,nikolov_sensitivityVSI_2022}.
Kinematic complexity is particularly important for calibrating materials to models that include nonlinear, anisotropic and/or co-dependent material parameters \cite{pierron_mt2_2021,pierron_mt2review_2023}.
Calibrating full-field data to complex material models is challenging in terms of designing a material specimen that will contain sufficient kinematic richness.
Designers must also consider experimental constraints, including boundary conditions and sample adhesion.
Current efforts to improve test design include the use of ``$\Sigma$-shaped'' specimens for anisotropic plastic parameter identification \cite{kim_pierronShapeOpt_2014} and topological optimization based on mechanical heterogeneity \cite{barroqueiro_topopt_2020,ihuaenyi_Mechanicsinformatics_2024,ihuaenyi_Mechanicsinformatics_2025}.
As experimental tests and demands become more complex, parameter extraction methods should leverage this richness.

\begin{figure}[t]
\centering
\includegraphics[width=170mm]{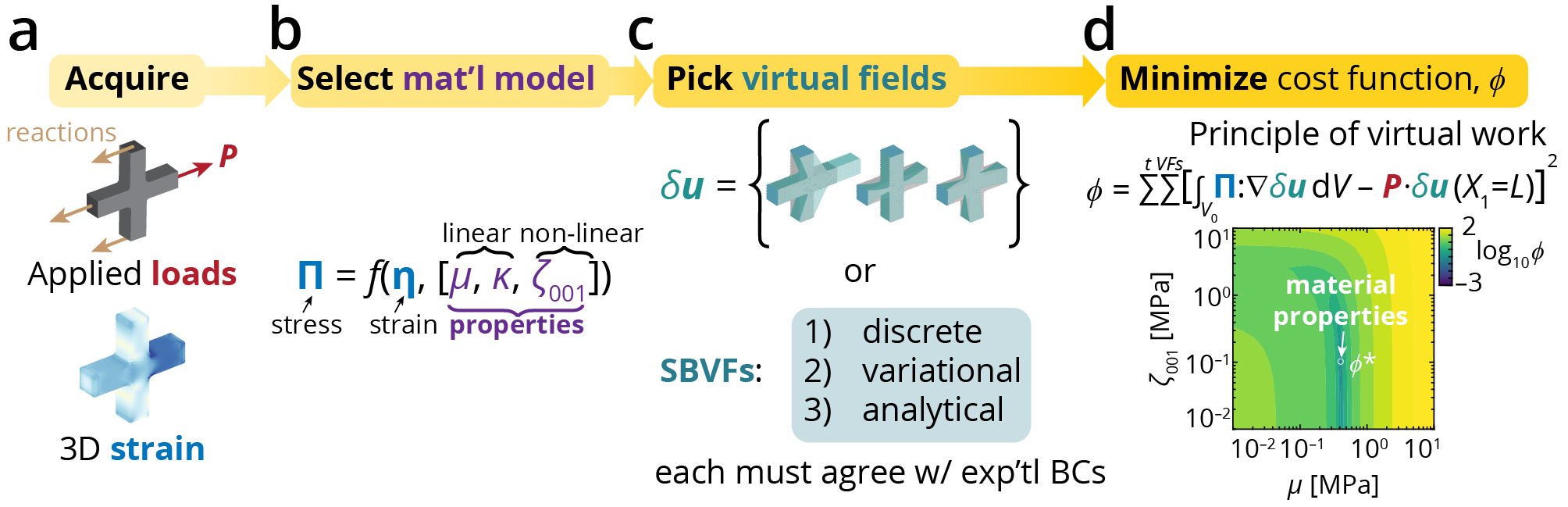}
\caption{\textbf{Virtual fields method flowchart for fixed choice of virtual fields.} (a) First, we measure our full-field data, (b) which is fitted to a material model. (c) The virtual fields method requires a selection of virtual fields for material parameter identification, which are defined by the user or an established procedure. (d) Fitting is performed via the principle of virtual work, and minimizing the resulting cost function space, $\phi$.}
\label{fig:VFM}
\end{figure}

\medskip
\noindent A different procedure from simple force-extension curve fitting is required to incorporate 2D or 3D full-field data into material characterization.
Techniques where model input parameters are estimated from comparison of model output magnitudes with experimental data (i.e., inverse methods) have been developed for full-field data measurements.
Common approaches include the virtual fields method (VFM) \cite{grediac_vfmAnisotropicPlates_1990,pagnacco_femu-fd_2007,pierron_vfmbook_2012}, finite element model updating (FEMU) \cite{kavanagh_femuorig_1971,cottin_femu-fd_1984,farhat_femu-d_1993}, and variational system identification (VSI) \cite{wang_VSIMRu_2021,nikolov_sensitivityVSI_2022}.
Both VFM and FEMU are used in characterizing anisotropic \cite{grediac_vfmAnisotropicPlates_1990,schmaltz_femuElastoplastic_2014}, viscoelastic \cite{giraudeau_viscoVFM_2005,giraudeau_viscoVFM_2006,marcot_viscoVFM_2023,tayeb_femuVisco_2023}, and biological materials \cite{avril_bioVFM_2010,rohan_bioFEMU_2013,romo_bioVFM_2014,bersi_bioVFM_2016,avril_bioVFM_2017}.
The main advantage of the VFM over FEMU is computational efficiency \cite{zhang_vfmvfemu_2017,kumar_femuvsVFM_2025} primarily due to the VFM directly calculating the stresses from the measured strains without a need to conduct forward simulations.
We show the VFM pipeline as a flowchart in \cref{fig:VFM}.
In cases where some kinematic data can be considered more reliable than other data, the manipulation of virtual fields further offers a straightforward, low-cost solution.

\medskip
\noindent The choice of the virtual fields plays a crucial role in the VFM.
In the absence of an automated procedure to define the virtual fields, user-defined virtual fields can be used \cite{pierron_vfmbook_2012}.
However, when extending the procedure to modeling non-linear behavior or calibrating material models with many parameters, we rely on the expertise of the investigator to select sufficiently good virtual fields \textit{sensitive} to BVPs' spatial regions of high kinematic richness.
Thus, the goal is to instead automatically select sensitivity-based virtual fields (SBVFs).
SBVF generation procedures have been initially developed for elasto-plastic material parameters \cite{avril_sensitivityVFMnoise_2004,marek_sensitivityVFM_2017} and consequently developed for anisotropic \cite{marek_sensitivityVFManiostropy_2019}, viscoelastic \cite{fletcher_IBIISensVFM_2021,fletcher_IBIISensVFM_2021a,matejunas_IBIISensVFM_2024} and hyperelastic parameters \cite{tayeb_hyperElastSensitivity_2019,tayeb_hyperElastSensitivity_2021}.
The present method to generate SBVFs involves evaluating the discrete difference in stress between two states of deformation, which may problematically amplify the effect of noise.
This problem is minimized with a high enough signal-to-noise ratio in the acquired data, but in cases of limited data acquisition, may not be universally acceptable.

\medskip
\noindent An additional choice in the VFM procedure is the selection of the material model and its respective material parameters.
Elastomeric mechanical behavior is classically calibrated to hyperelastic models \cite{holzapfel_nonlin_mech_2000}.
For example, the Mooney--Rivlin model \cite{mooney_mooneymodel_1940} and the Yeoh model \cite{yeoh_yeohmodel_1989,yeoh_yeohmodel_1993} are broadly used for characterizing elastomers especially by simultaneously incorporating data in uniaxial, biaxial and shear strain states.
Unfortunately, traditional hyperelastic models involve covariant kinematic terms (the invariants of the left or right Cauchy-Green tensors) making it difficult for invariant-based models to distinguish between model term contributions (see \cref{sec:theory:hyperelastic} for more details).
Thus a solution is to either calibrate data from multiple boundary value problems simultaneously or incorporate a material model that decouples these effects beforehand.
J. Criscione et al. introduced a constitutive formulation using decomposed kinematics via logarithmic strain-based invariants~\cite{criscione_basickinv_2000}.
Henceforth we refer to this family of models as natural strain invariant models.
Further developments of the natural strain invariant model extend to anisotropic \cite{criscione_orthkinv_2001} and elastomeric foam characterization \cite{landauer_kinvFoams_2019,li_foamKinv_2022}.

\medskip
\noindent The goal of this work is to design a pipeline that leverages full three-dimensional kinematic data, models based on strain invariants, and an improvement on the SBVF method to quantifiably improve material parameter calibration.
Demonstrating the improvement over our prior work \cite{estrada_MRu_2020}, we fit a set of low signal strain data to both the Mooney-Rivlin model and the natural strain invariant model comparing parameter convergence between user-defined and sensitivity-based virtual fields.
We then quantify the robustness of the technique for experimental data and illustrate how we could improve material model calibration by modifying the sample shape and loading conditions.

\section{Theory and Methods}
\label{sec:theory}

We begin with a material sample occupying a volumetric region $\Omega_0$ in its reference configuration.
A point in the sample is described by internal coordinates $\bX \in \Omega_0$.
At a later time $t$ during a test, the sample occupies a region $\Omega$ and a material point $\bX$ occupies a spatial coordinate $\bx \in \Omega$.
The reference and deformed coordinates are related through a deformation mapping $\deformmap$, such that $\bx = \deformmap\left(\bX,t\right)$.
Accordingly, we define the displacement field $\bu(\bX, t)=\bx-\bX$.
The local deformation at a material point is described by the deformation gradient tensor, $\up{F}(\bX,t)$,
\begin{equation}
\up{F} =
\bnabla_{\bX} \deformmap
= \bnabla_{\bX} \bu + \up{I} \, ,
\end{equation}
where $\up{I}$ is the second-order identity tensor and $\bnabla_{\bX}(\cdot)$ is the material gradient operator.
The Jacobian determinant,
$J(\bX,t)=\det(\up{F}(\bX,t))$, represents the local volumetric expansion.
\medskip

\noindent
We partition the boundary of the sample in the reference configuration, $\partial\Omega_0$, into disjoint parts,
such that $\partial\Omega_0 = \partial\Omega_0^t \cup \Omega_0^u$ and $\Omega_0^t \cap \Omega_0^u = \emptyset$.
The traction-bearing part of the sample surface, $\partial \Omega_0^t$, is subjected to applied tractions $\bt(\bX,t) = \overline{\bt}$
and the displacement-prescribed part of the sample surface, $\partial \Omega_0^u$, is subjected to displacements $\bu(\bX,t) = \overline{\bu}$.
In practice, the displacement of a sample edge, e.g., $u_1(X_1 = L)$, is often prescribed while the tractions are only known in an aggregate sense, i.e.,
\begin{equation} \label{eqn:load}
\bP=\int_{\partial\Omega_0^t} \bt ~dA.
\end{equation}
The displacement field in the sample, $\bu(\bX,t)$, is sampled discretely on a set of reference measurement or simulation points.

\subsection{Principle of virtual work and the virtual fields method}

The virtual fields method (VFM) leverages the principle of virtual work (PVW) with full-field kinematic and global kinetic data to inversely calibrate material models.
Generally, the VFM will match the internal and external virtual work by tuning the parameters $\bm\xi$ of a chosen material model $\mathcal{M}$, given an experimentally measured displacement field $\bu(\bX)$ and the global force data $\bP$.
As our experimental kinematic data is typically acquired in the reference configuration while our load cell information acquires current force measurements \cite{estrada_MRu_2020, nikolov_sensitivityVSI_2022}, we write the PVW as a function of the first Piola-Kirchhoff stress $\bm{\Uppi}(\up{F},\bm{\xi})$,
\begin{equation} \label{eq:pvw}
\displaystyle \int_{\Omega_0} \bm{\Uppi}:(\bm{\nabla}_{\bm{X}} \delta\bu) ~ dV_0 - \int_{\partial {\Omega_0}} (\bm{\Uppi} \bm{N}) \cdot \delta\bm{u} ~dA_0 = 0,
\end{equation}
where $\bm{N}$ is the outward surface normal to $\partial\Omega_0$ and $\delta \bu(\bX)$ represents a small, ``virtual'' perturbation on the displacement field $\bu(\bX)$.
According to Cauchy's stress theorem, a point on $\partial\Omega_0$ has traction $\bt = \Uppi\bm{N}$.
The virtual field $\delta \bu(\bX)$ has the only restriction of requiring the same boundary condition types on the surfaces of the sample, and not the values specifically---i.e. if a sample is extended along its length, a valid choice of virtual field may have a zero, or longer extension.
We direct the interested reader to a discussion on virtual field restrictions in \cite{pierron_vfmbook_2012}.

\medskip
\noindent Considering the boundary data typical in our own experiments---extension of one surface, described by $X_1 = L$, in the first Cartesian direction $\up{\bm{e}}_1$ and a measured force $P$---we write the principle of virtual work as
\begin{equation}
\label{eq:vfm-specific}
\int_{\Omega_0} \bm{\Uppi}:(\bm{\nabla}_{\bm{X}} \delta \bu)~ dV_0 - P \cdot \delta u_1(X_1=L) = 0.
\end{equation}

\medskip\noindent
The virtual fields method then takes multiple experimental data sets $\{\up{F}(\bX,t), P(t)\}$ and virtual fields $\delta \bm{u}(\bX)$ and simultaneously minimizes the sum of several instances of \cref{eq:vfm-specific} as a cost function $\phi$ by varying guesses of material parameters $\bm{\xi}$ in $\bm{\Uppi}(\up{F},\bm{\xi})$, i.e.,
\begin{equation}
\label{eqn:vfm-PK}
\phi \equiv \sum\limits_{i=1}^{N_{\rm exp}} \sum\limits_{j=1}^{n_{\rm VF}} \left(\int_{\Omega_0} \bm{\Uppi}^{(i)}:(\bm{\nabla}_{\bm{X}} \delta\bm{u}^{(j)}) dV_0 - P^{(i)} \cdot \delta u_1^{(j)}(X_1=L) \right)^2,
\end{equation}
where $N_{\rm exp}$ represents the number of experiments carried out and $n_{\rm VF}$ represents the total number of virtual fields considered.
The output of the VFM is then the best estimate of material properties $\bm{\xi}^*$ for a particular model $\mathcal{M}$, which is determined as
\begin{equation}
\bm{\xi}^* = \argmin_{\bm{\xi}} (\phi).
\end{equation}

\subsection{Hyperelastic material models and invariants of deformation} \label{sec:theory:hyperelastic}

Herein we consider hyperelastic models with additively decoupled isochoric strain energy density functions $\psi_{\rm iso}$ and volumetric response functions $\psi_{\rm vol}$ for our material model calibration.
Restricting our analysis to isotropic materials, we may utilize the representation theorem for invariants~\cite{holzapfel_nonlin_mech_2000} and express strain energies
as functions of the invariants $I_i$ (where $i=1, 2, 3$) of the left and right Cauchy-Green deformation tensors, $\up{B}(\bX) = \up{F} \up{F}^\intercal $ and $\up{C}(\bX) = \up{F}^\intercal \up{F}$, respectively.
However, the functions $I_i$ cannot be completely decoupled from one-another---$I_1$ and $I_2$ are covariant via the underlying stretch state, even when put into their isochoric forms $\bar{I}_1 = J^{-2/3} I_1$ and $\bar{I}_2 = J^{-4/3} I_2$---which is limiting from an experimental standpoint.
If a model depending on, say, $I_1$ and $I_2$ is to be calibrated, an ideal experiment would be capable of varying $I_1$ independently from $I_2$, and vice versa.

\begin{figure}[t]
\centering
\includegraphics[width=85mm]{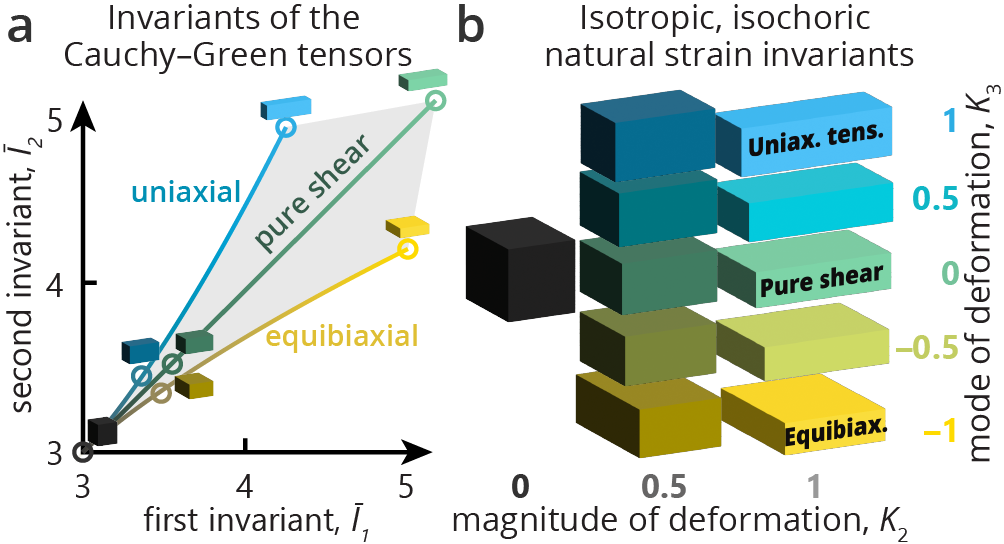}
\caption{\textbf{Invariants of deformation tensors.} (a) Illustration fo non-orthogonal isochoric invariants $\bar{I}_1$ and $\bar{I}_2$ of the Cauchy-Green tensors $\mathbf{B}$ and $\mathbf{C}$ as functions of principal stretches. (b) Visualization of orthogonal, physically meaningful invariants, representing independent modes and magnitudes of deformation.}
\label{fig:orthDecomp}
\end{figure}

\medskip
\noindent
The problem of the $I_1-I_2$ covariance can be solved by using alternative invariants of strain, instead of those pertaining to $\up{C}$ and $\up{B}$.
This is shown schematically in \cref{fig:orthDecomp}.
The reformulation of both isotropic \cite{criscione_basickinv_2000} and orthotropic \cite{criscione_orthkinv_2001} hyperelastic models was developed by J. Criscione et al. for this purpose.
Instead of $I_i(\up{C})$, three invariants $K_i$ of logarithmic (Hencky) strain $\bm{\upeta}(\bX)$ are presented, where
\begin{subequations}
\label{eq:natinvars}
\begin{align}
K_1&=\text{tr}\left(\bm{\upeta}\right)=\ln{J},\\
K_2&=\left|\text{dev}\left(\bm{\upeta}\right)\right|=\sqrt{\text{dev}\left(\bm{\upeta}\right):\text{dev}\left(\bm{\upeta}\right)},\\
K_3&=3\sqrt{6}\;\text{det}\left(\bm{\Upphi}\right),
\end{align}
\end{subequations}
where tr($\cdot$) represents the trace and dev($\cdot$) represents the deviatoric part of a quantity, the (left) logarithmic strain is given by
\begin{equation}
\bm{\upeta}=\frac{1}{2}\log(\up{B}),
\end{equation}
and the tensor $\bm{\Upphi}=\text{dev}\left(\bm{\upeta}\right)/K_2$, introduced by \textcite{criscione_basickinv_2000} represents the tensorial direction of deviatoric strain.
Logarithmic strain has an added experimental design benefit in that the domain of values is symmetric and spans $(-\infty,\infty)$ in contrast to, e.g., Green--Lagrange strain $(-\frac{1}{2},\infty)$ or Euler--Almansi strain $(-\infty, \frac{1}{2})$.
The natural strain invariants $K_i$ in \cref{eq:natinvars} represent the amount of dilatation, magnitude of distortion, and the mode of distortion, respectively.

\medskip
\noindent
A general polynomial series for the total isotropic, hyperelastic strain energy density function $\psi$ was developed and presented by Criscione in \cite{criscione_basickinv_2000},
\begin{equation}
\label{eq:gencrisc}
\psi = \sum_{i=1}^\infty \alpha_i K_1^i + K_2^2 \left(\mu + \sum_{i=1}^\infty \beta_i K_1^i \right) + K_2^3 \sum_{i=0}^\infty \sum_{j=0}^\infty \sum_{k=0}^\infty \zeta_{ijk} K_1^i K_2^j K_3^k,
\end{equation}
where $\alpha_1$ represents the reference configuration pressure, the remaining $\alpha_i$ are material constants describing the pure bulk behavior, $\mu$ is the shear modulus, $\beta_i$ describe coupled bulk and isochoric deformations, and $\zeta_{ijk}$ describe material behavior that may depend simultaneously on the magnitude and mode of deformation.
For isotropic, nearly incompressible hyperelastic materials, a subset of the terms introduced in \cref{eq:gencrisc} suffice to describe the isochoric material response,
\begin{equation} \label{eqn:crisland}
\psi_{\rm iso}=\mu K_2^2+\zeta_{010}K_2^4+\zeta_{001}K_3K_2^3.
\end{equation}
This strain energy density function builds on a closely related form for hyperelastic foams presented by \textcite{landauer_kinvFoams_2019}.
The models $\mathcal{M}$ compared in the study are summarized with their parameters $\bm{\xi}$ in \cref{tab:models}.

\medskip
\noindent
The materials are assumed to behave with limited compressibility, such that the energy storage due to bulk deformation can be additively composed into
the isochoric part $\psi_{\rm iso}$ and the volumetric part $\psi_{\rm vol}$, described by a quadratic term, i.e.,
\begin{equation} \label{eq:vol_strain_energy}
\psi_{\rm vol}(K_1) = \frac{\kappa}{2}K_1^2,
\end{equation}
where $\kappa$ is the bulk modulus.

\begin{table}[t]
\centering
\begin{tabular}{l l l}
\toprule
\textbf{Model, $\mathcal{M}$} & \textbf{Model parameters, $\bm{\xi}$} & \textbf{Isochoric strain energy, $\psi_{\rm iso}$} \\
\midrule
neo-Hookean & $\left[\mu\right]$ & $\frac{\mu}{2}\left(\bar{I}_1-3\right)$ \\
Mooney-Rivlin & $\left[C_{10},C_{01}\right]$ & $C_{10}\left(\bar{I}_1-3\right)+C_{01}\left(\bar{I}_2-3\right)$ \\
Mooney-Rivlin (mixture) & $\left[\mu,\alpha\right]$ & $\frac{\mu(1-\alpha)}{2}\left(\bar{I}_1-3\right)+\frac{\mu \alpha}{2}\left(\bar{I}_2-3\right)$\\
isotropic natural strain & $\left[\mu,\zeta_{010},\zeta_{001} \right]$ & $\mu K_2^2+\zeta_{010}K_2^4+\zeta_{001}K_3K_2^3$\\
\bottomrule
\end{tabular}
\caption{Isochoric parts of the strain energy density functions used in the present study.}
\label{tab:models}
\end{table}

\subsection{Sensitivity-based virtual fields} \label{sec:theory:sbvf}

The VFM for our experimental setup as in \cref{eq:vfm-specific} requires us to acquire a displacement field $\bu(\bX)$ and a global load $P$, as well as a choice of material model.
While \cref{eq:vfm-specific} will hold for any valid virtual field $\delta \bu(\bX)$, some fields provide more information than others.
In this section, we extend the (discrete) sensitivity-based virtual fields (SBVF) developed by \textcite{marek_sensitivityVFM_2017} which creates optimized virtual fields for the calibration of material models.

\subsubsection{Discrete approach (dSBVFs)}
We first briefly present the discrete sensitivity-based virtual fields method.
User-defined virtual fields (UDVFs) aim to probe the mechanical response of a material subjected to boundary conditions and a known kinematic field.
However, some choices of virtual field are better suited to this purpose than others.
The natural extension, therefore, is optimizing the virtual fields themselves such that each material model parameter of interest is identified with a best, or most \textit{sensitive}, virtual field, given experimental data $\{\bu(\bX,t), P(t)\}$.
Sensitivity-based virtual fields (SBVF) were originally formulated by \textcite{marek_sensitivityVFM_2017} for elasto-plastic material parameter identification and later adapted for hyperelasticity \cite{tayeb_hyperElastSensitivity_2019,tayeb_hyperElastSensitivity_2021}.
The sensitivity of the first Piola-Kirchhoff stress $\bm{\Uppi}(\up{E}, \bm{\xi})$---where $\up{E}(\bm{X},t)$ represents the strain field calculated from $\bu(\bX,t)$ and $\bm{\xi}$ represents a current guess of material properties in the VFM---to both a change in a material quantity and a deformation state is defined discretely using two differentials.
First, the change in stress $\delta \bm{\Uppi}^{(j)}$ from a perturbation in the guess of the $j^{\rm th}$ material property $\xi_j$ is
\begin{equation}
\delta\bm{\Uppi}^{(j)}\left(\bm{\xi}, t\right) = \bm{\Uppi}\left(\bm{\xi}+\delta\xi_j,t\right) - \bm{\Uppi}\left(\bm{\xi},t\right),
\end{equation}
where the chosen material model $\mathcal{M}$ is parameterized by the set of material parameters $\bm{\xi}$.

\medskip
\noindent The second differential is evaluated between deformation states $\up{E}(t)$ parameterized by time steps $\Delta t$.
The stress sensitivity $\delta\tilde{\bm{\Uppi}}^{(j,k)}(\up{E}(t_k),\bm{\xi})$ at a time $t_k$ is then defined as
\begin{equation} \label{eqn:linear_sens}
\delta\tilde{\bm{\Uppi}}^{(j,k)} = \delta\bm{\Uppi}^{(j)}\left(t_k \right) - \delta\bm{\Uppi}^{(j)}\left(t_k-\Delta t\right).
\end{equation}
Comparatively large values of $\delta\tilde{\bm{\Uppi}}^{(j,k)}$ imply a large sensitivity of the stress output on changes of the value of a material parameter given a current kinematic field.

\medskip
\noindent The stress differentials are then converted into virtual displacement fields that adhere to the form of the boundary conditions of the calibration experiments.
In practice, boundary condition adherence is achieved using the shape functions of a finite element mesh via
\begin{equation} \label{eqn:Bmat_bar}
\delta\bu^{(j,k)}(t_k)=\text{pinv}(\bar{\up{B}})\delta\tilde{\bm{\Uppi}}^{(j,k)}\left(\bm{\xi},t_k\right),
\end{equation}
where $\text{pinv}(\cdot)$ designates the pseudo-inverse operator and $\bar{\up{B}}$ is the modified global strain-displacement matrix with prescribed displacement boundary conditions enforced onto the original global strain-displacement matrix $\up{B}$.
The virtual displacement gradient field used in the principle of virtual work is then calculated as
\begin{equation} \label{eqn:Bmat}
\nabla_{\bX}\delta\bu^{(j,k)}(t_k)=\up{B}\delta\bu^{(j,k)}(t_k).
\end{equation}

\subsubsection{Variation-based approach (vSBVFs)}
\label{sec:SBVFvarbased}
\noindent
A new alternative, but philosophically related, approach to the discrete method above is one wherein we equate the variation of stress with respect to a material parameter of interest with the variation of stress with respect to displacement.
By setting the two variations equal, we can analytically solve for the virtual (i.e., variational) displacement field that directly corresponds to the variation of a specific parameter.
To find the (first) variation of a spatial-configuration quantity $\quant$ along a parameter $\alpha$, we use the directional G\^ateaux derivative~\cite{holzapfel_nonlin_mech_2000},
\begin{equation}
\delta_{\alpha}\quant = \fracd{}{\epsilon} \left(\quant[\alpha + \epsilon \, \delta\alpha] \right) \Big|_{\epsilon = 0}.
\end{equation}

\noindent For the variation-matching procedure, we use the rotated Kirchhoff stress
\begin{equation}
\rotkir = J \, \up{R}^{\transpose}\cauchy\up{R} \, ,
\end{equation}
where $\cauchy = J^{-1}\pkone\up{F}^{\transpose}$ is the Cauchy stress and $\up{R}$ is the rotation tensor arising from the polar decomposition, $\up{F} = \up{R}\up{U}$.
We show in appendix \cref{sec:app:variation_approach_solution} that the use of $\rotkir$ is crucial to the success of vSBVFs.

\noindent
Suppose we cast the variation of $\rotkir$ for both an infinitesimal variation of a single material parameter $\delta_{\xi_j}$ and a corresponding displacement variation $\delta_{\bu} = \delta_{\bu_j}$ such that $\delta_{\bu}
\rotkir = \delta_{\xi_j} \rotkir$, or
\begin{equation}
\label{eq:matchVariations}
\fracd{}{\epsilon} \left(\rotkir[\bu + \epsilon \delta\bu
, \bm{\xi}] \right) \Big|_{\epsilon = 0} = \fracd{}{\epsilon} \left(\rotkir[\bu, \bm{\xi}[\xi_j + \epsilon \delta\xi_j] \right) \Big|_{\epsilon = 0}.
\end{equation}
Our goal then becomes to carry out the variations of $\rotkir$ on the left and right sides of \cref{eq:matchVariations} and then determine the displacement variation $\delta\bu_j$---i.e., one new sensitivity-based virtual displacement field---in terms of a material property variation $\delta\xi_j$.

\medskip
\noindent Whereas $\delta_{\xi_j} \rotkir$
is relatively straightforward to acquire, $\delta_{\bu} \rotkir$ is more unwieldy in general.
For a hyperelastic material, we can write the relationship between the stress variation and a corresponding strain variation $\delta_{\bu}\up{E}$ using a fourth-order stiffness tensor $\stiffA(\bm{\xi}, \bu)$ as
\begin{align}
\label{eq:stiffness_tensor_defined}
\begin{aligned}
\delta_{\bu}
\rotkir &= \stiffA \, \delta_{\bu}
\up{E} \, , \\
\left(\delta_{\bu}\rotkir\right)_{ij} &= \stiffA_{ijkl} \left( \delta_{\bu}\up{E} \right)_{kl} \, .
\end{aligned}
\end{align}
To find the virtual strain $\delta_{\bu}\up{E}$ corresponding to $\delta_{\bu} \rotkir = \delta_{\xi_j} \rotkir$, we invert the stiffness tensor $\stiffA$ to evaluate
\begin{equation}
\delta_{\bu}\up{E} \left( \delta\bu = \delta\bu_j \right) =\stiffA^{-1} \,\delta_{\xi_j}\rotkir.
\label{eq:vSBVF_inversion}
\end{equation}

\medskip
\noindent
We note that a solution of the virtual strain $\delta_{\bu}\up{E}$ does not correspond to a unique choice of $\delta_{\bu}$.
In the case where $\up{E} = \up{E}_{\rm G} = \left(\up{C}-\up{I}\right)/2$, the Green--Lagrange strain, its variation is related to that of the deformation gradient $\up{F}$ as
\begin{equation}
\delta_{\bu} {\up{E}}_{\rm G} = \sym{ {\up{F}}^{\transpose} \, \delta_{\bu} {\up{F}} },
\end{equation}
in which $\delta_{\bu}\up{F} = \nabla_{\bX} \varu$.
To proceed, we elect to require that the virtual displacement field satisfies
\begin{equation}
{\up{F}}^{\transpose} ~\delta_{\bu}{\up{F}} = \sym{ {\up{F}}^{\transpose} ~\delta_{\bu}{\up{F}} }.
\label{eq:no_virtual_rotation_assumption}
\end{equation}
This leads to a unique solution
\begin{equation}
\nabla_{\bX}\delta\bu_j=\up{F}^{-\intercal} \stiffA^{-1}
\, \delta_{\xi_j}\rotkir \, .
\end{equation}
We now follow the same procedure in \cref{eqn:Bmat_bar,eqn:Bmat} to satisfy kinematic compatibility between $\nabla_{\bX}\delta \bu_j$ and $\delta \bu_j$.
The benefit of this technique is that it procedurally defines virtual displacement gradient fields for any set of material model parameters with an explicit evaluation of virtual strain fields.
The general procedure for finding the virtual field $\nabla_{\bX}\delta_{\xi_j}\bu$ specifically for the Mooney-Rivlin material model and the natural strain invariant model parameters are highlighted in appendix \cref{sec:app:variation_approach_solution}.

\subsubsection{Analytical derivative approach for natural-invariant-type models (aSBVFs)}
\label{sec:SBVFderivative}
\noindent In our prior work \cite{nikolov_sensitivityVSI_2022} we presented an derivative-based method for SBVFs based on a set of stretch-based invariants.
We cast this method in terms of the natural strain invariants $K_i$ here for comparison to the two previous methods.
Given a measured displacement field $\bu(\bX)$ converted into a Hencky strain field $\bm{\upeta}(\bu(\bX))$, material model parameters $\bm{\xi}$, and strain energy density function $\psi(K_1, K_2, K_3)$, the virtual Hencky strain fields $\delta\bm{\upeta}(\bu(\bX))$ at an experimental timestep $t_i$ are given as
\begin{equation} \label{eqn:sbvf-deriv}
\delta\bm{\upeta}^{(i,j,k)}=\frac{\partial^3\psi(K_1^{(k)},K_2^{(k)},K_3^{(k)})}{\partial\bm{\upeta}\partial K_i\partial\xi_j}.
\end{equation}
We note that $\bm{\uptau}(\bX) = \partial \psi / \partial \bm{\upeta}(\bX)$ is the Kirchhoff stress tensor.
For an isotropic solid, $\bm{\uptau}$ and $\bm{\upeta}$ are work conjugates \cite{hoger_henckyConjugate_1987}.

\begin{figure}[p]
\centering
\includegraphics[width=170mm]{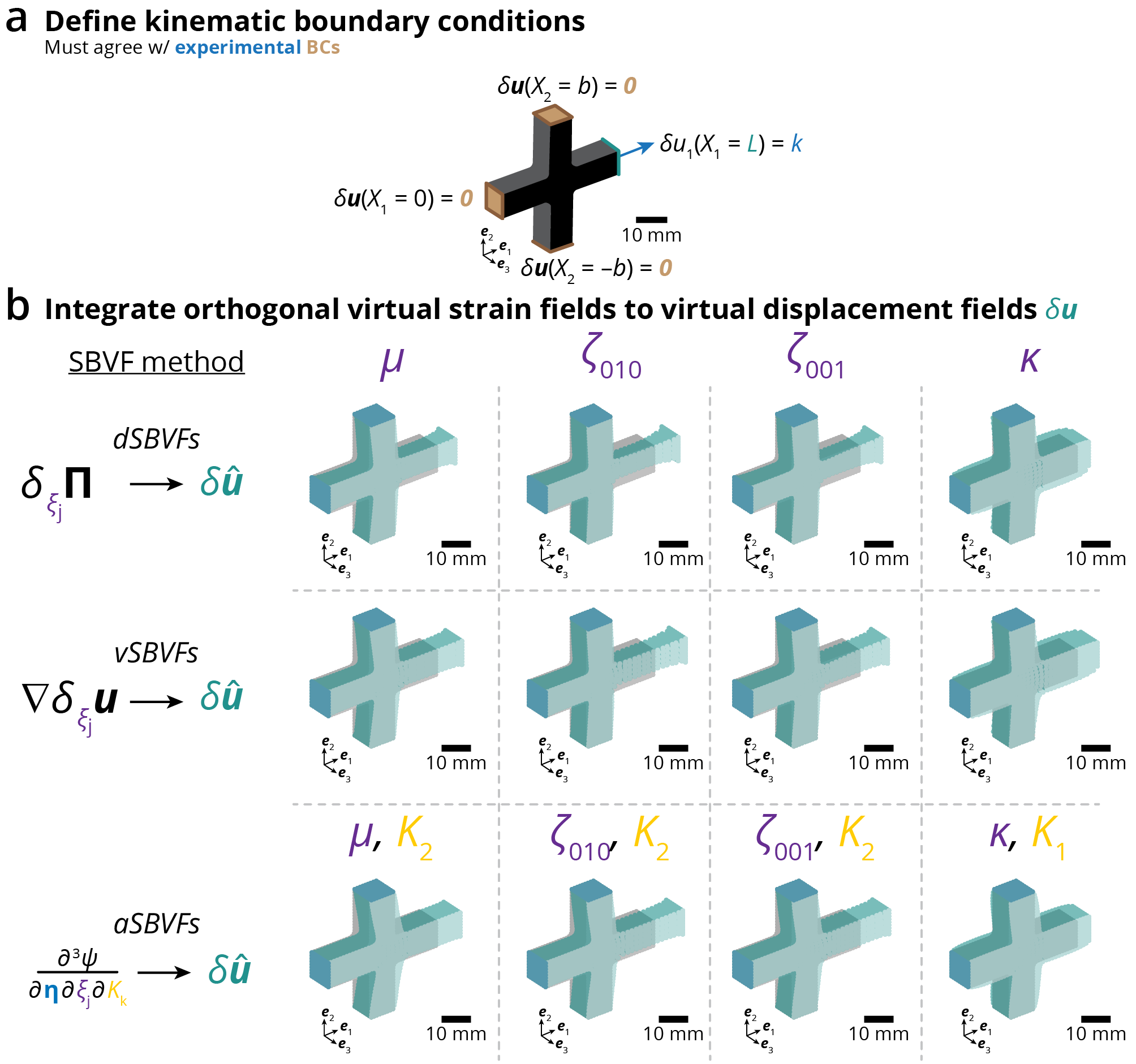}
\caption{\textbf{Representative virtual fields.} (a) Boundary conditions imposed on the sample, with displacements fixed (brown) or prescribed (blue). (b) Numerically integrated virtual displacement fields corresponding to each material parameter, for SBVFs representing $\mu$, $\zeta_{010}$, $\zeta_{001}$, and $\kappa$.}
\label{fig:fields}
\end{figure}

\medskip
\noindent
As work conjugate quantities for our material class of interest, we can adapt the PVW in \cref{eq:vfm-specific} and incorporate the experimental boundary conditions as
\begin{equation}
\label{eq:vfm-hencky}
\int_{\Omega} \bm{\uptau}:\delta \bm{\upeta}~ dV - P \cdot \delta u_1(X_1=L) = 0.
\end{equation}
From $\delta \bm{\upeta}$ we need to determine the displacement $\delta \bu$; however, all strain metrics remove the skew (i.e. rotation) information.
Thus, we use the point-wise rotation tensor $\up{R}(\bX)$ from the left polar decomposition encoded in the experimental data to convert from virtual stretch to virtual displacement gradient as,
\begin{equation}
\label{eq:DeltaGraduToDeltaV}
\nabla_{\bX}\delta \bu = (\delta\up{V}) \up{R} - \up{I},
\end{equation}
where the left stretch tensor $\up{V} = \sqrt{\up{B}}$.
As the virtual stretch and virtual strain are related through
\begin{equation}
\label{eq:DeltaEtaToDeltaV}
\delta\bm{\upeta}=\log(\delta\up{V}),
\end{equation}
we thus combine \cref{eq:DeltaGraduToDeltaV,eq:DeltaEtaToDeltaV,eqn:Bmat} to ensure consistency between each individual $\delta \bm{\upeta}^{(i,j,k)}$ and its corresponding $\delta \bu^{(i,j,k)}$.
A set of representative virtual fields for dSBVFs, vSBVFs, and aSBVFs are shown in \cref{fig:fields}.

\subsection{Computational Approach to Elasotmeric Material Characterization} \label{sec:methods:comput}

\noindent Our new SBVFM method of \cref{sec:SBVFvarbased,sec:SBVFderivative} was developed and validated using finite element simulations with chosen material geometries, models, properties, and boundary conditions.
These parameters are initialized in an input file in MATLAB (The Mathworks, Natick, MA).
For all simulations, we used material parameters corresponding to Treloar's data from vulcanized rubber, i.e. $\bm{\xi}=\left[\mu,\alpha,\kappa\right]=\left[\SI{405}{\kilo\pascal},0.037,\SI{3.915}{\mega\pascal}\right]$ for stretch invariant models and $\bm{\xi}=\left[\mu,\zeta_{010},\zeta_{001},\kappa\right]=\left[\SI{405}{\kilo\pascal},\SI{2.50}{\kilo\pascal},\SI{102}{\kilo\pascal},\SI{3.915}{\mega\pascal}\right]$ for the simplified isotropic natural strain invariant model.
Virtual pull tests were performed at $\max(u_1(\bX))=[2,4,8]\,\SI{}{\mm}$, and forward runs of the boundary value problem were performed in Abaqus/Standard (SIMULIA, Dassault Syst\'emes) at object (\SI{40}{\mm} length) and mesh ($\sim$\SI{0.5}{\mm}) sizes approximating our experimental capabilities \cite{estrada_MRu_2020, scheven_physicsMRu_2020}.

\section{Results}
Herein we illustrate (a) the benefit of SBVFs over UDVFs, (b) the successful convergence of all SBVF methods, (c) the comparison of convergence across boundary value problems (BVPs) of different kinematic complexity, and (d) the numerical effect of noise in the displacement gradient on the convergence of material properties.

\begin{figure}[t]
\centering
\includegraphics[width=170mm]{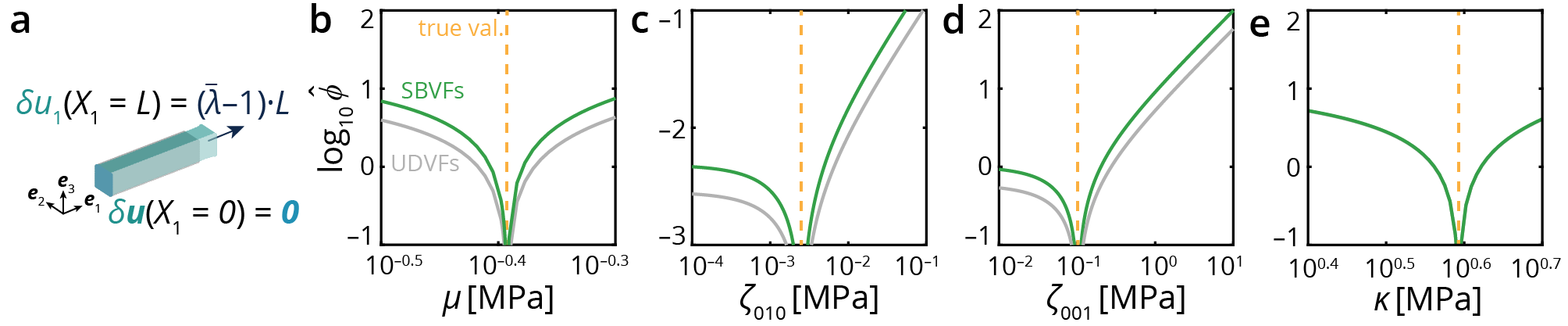}
\caption{\textbf{Comparison of cost function convergence for user-defined (grey) and sensitivity-based virtual fields (green).} (a) A rectangular prism of dimensions 40$\times$8$\times$\SI{8}{\mm} and properties $\bm{\xi}=\left[\mu,\zeta_{010},\zeta_{001},\kappa\right]=$[\SI{405}{\kilo\pascal},\SI{2.50}{\kilo\pascal},\SI{102}{\kilo\pascal},\SI{3.915}{\mega\pascal}] has encastred distal faces and is stretched to a set displacement of \SI{8}{\mm}. (b--e) Normalized cost function profiels near the true material parameter values are shown for (b) shear moduslus $\mu$, (c) non-linear stiffening modulus $\zeta_{010}$, (d) mode-dependent modulus $\zeta_{001}$, and (e) bulk modulus $\kappa$, comparing user-defined (gray) and SBVF (green) virtual fields.}
\label{fig:gen_v_sbvf}
\end{figure}

\medskip
\noindent
To compare parameter convergence across VFM methodologies, we (1) modify the cost function $\phi$ to correct for the work done on the sample and (2) adjust, or ``equalize'' each virtual field $\delta \bu$ such that each contributes a similar amount of energy \cite{estrada_MRu_2020}.
The modified cost function $\hat\phi$ is thus
\begin{equation}
\hat{\phi} \equiv \sum\limits_{i=1}^{N_{\rm exp}} \sum\limits_{j=1}^{n_{\rm VF}} \left(\frac{\int_{\Omega_0} \bm{\Uppi}^{(i)}:(\bm{\nabla}_{\bm{X}} \delta\bm{\hat{u}}^{(j)}) dV - P \cdot \delta \hat{u}_1^{(j)}(X_1=L) }{P^{(i)} \cdot u_1^{(i)}(X_1=L)}\right)^2,
\end{equation}
where $\delta \bm{\hat{u}}^{(j)}$ represent the virtual displacement fields equalized with respect to the $l^2$-norms of the virtual field gradients $\text{mean}(||\bm{\nabla}_{\bm{X}}\delta\bu(\bm{X})||)$.
\Cref{fig:gen_v_sbvf} shows $\hat{\phi}$ the behavior of the normalized cost function $\hat{\phi}$ around the best-fit parameters $\bm{\xi}^*$ (orange dashed line).
To quantify $\hat{\phi}$-space specificity, we calculate the local sharpness via one-sided finite differences in the neighborhood of $\bm{\xi}^*$.
The cost function sharpness $\varsigma_{\bm{\xi}_i}(\bm{\xi}^*)$ is defined, in a manner similar to Keskar et al. \cite{keskar_sharpness_2017}, as
\begin{equation}
\label{eq:sharpness}
\varsigma_{\xi_i}(\bm{\xi}^*) = \lim_{\epsilon\rightarrow 0} \frac{\max(\phi(\xi_i^*\pm\epsilon)) - \phi(\xi_i^*)}{1+\phi(\xi_i^*)}.
\end{equation}
Cost function sharpness values along different material parameters were found for a variety of choices of boundary value problems, macroscale stretches, and virtual fields in \cref{tab:sharpness}.

\medskip
\noindent The benefit of SBVFs in the absence of noise, regardless of the exact SBVF form (all three types overlap on the scales of \cref{fig:gen_v_sbvf}) is apparent from the comparatively sharper green vs. gray peaks in panels (b-e).
The values for sharpness were consistently $\sim2-3\times$ larger for SBVFs as compared to UDVFs for both the Mooney-Rivlin models (traditional and mixture-parameter) and the natural strain invariant model.

\begin{table}[t]
\centering
\small
\begin{tabular}{l l l|l l l|l l l|l l l l}
\toprule
& & & \multicolumn{3}{c}{\textbf{M-R}} & \multicolumn{3}{c}{\textbf{M-R (mixture)}} & \multicolumn{4}{c}{\textbf{Isotropic natural strain}} \\
\midrule
Sample geom. & $\bar{\lambda_1}$ & VFs & $\varsigma_{C_{10}}$ & $\varsigma_{C_{01}}$ & $\varsigma_\kappa$ & $\varsigma_\mu$ & $\varsigma_\alpha$ & $\varsigma_\kappa$ & $\varsigma_\mu$ & $\varsigma_{\zeta_{010}}$ & $\varsigma_{\zeta_{001}}$ & $\varsigma_\kappa$ \\
\midrule
rect. uniax. & 1.2 & UDVF & 91.81 & 86.14 & 3.688 & 45.80 & 1.285 & 3.688 & 44.53 & 1.136 & 5.836 & 3.682 \\
& & dSBVF & 131.4 & 123.4 & 3.782 & 65.57 & 1.832 & 3.782 & 78.28 & 2.006 & 10.24 & 3.783 \\
& & vSBVF & 129.4 & 121.5 & 3.707 & 64.53 & 1.767 & 3.707 & 77.72 & 1.937 & 10.06 & 3.721 \\
& & aSBVF & -- & -- & -- & -- & -- & -- & 81.39 & 2.135 & 10.80 & 3.649 \\
\midrule
rect. uniax. & 2 & UDVF & 21.98 & 16.44 & 0.9422 & 10.88 & 1.225 & 0.9422 & 9.559 & 4.387 & 5.636 & 0.9439 \\
& & dSBVF & 32.45 & 24.18 & 1.143 & 16.07 & 1.834 & 1.143 & 17.11 & 8.163 & 10.24 & 1.202 \\
& & vSBVF & 26.50 & 20.35 & 0.7477 & 13.132 & 1.331 & 0.7477 & 15.27 & 5.601 & 8.153 & 0.8483 \\
& & aSBVF & -- & -- & -- & -- & -- & -- & 18.62 & 8.604 & 11.04 & 0.8995 \\
\midrule
mod. biax. & 1.2 & UDVF & 89.47 & 84.33 & 3.643 & 44.64 & 1.180 & 3.643 & 43.46 & 1.113 & 5.362 & 3.645 \\
& & dSBVF & 160.2 & 150.4 & 4.993 & 79.91 & 2.240 & 4.993 & 101.3 & 3.575 & 14.65 & 5.060 \\
& & vSBVF & 157.6 & 147.9 & 4.709 & 78.61 & 2.207 & 4.709 & 100.4 & 3.437 & 14.36 & 4.812 \\
& & aSBVF & -- & -- & -- & -- & -- & -- & 103.5 & 3.278 & 14.11 & 4.068 \\
\bottomrule
\normalsize
\end{tabular}
\caption{Cost function sharpness across choices of boundary conditions, average stretch, and virtual field type. In all cases, properties in forward simulations were set to $\bm{\xi}=\left[\mu,\zeta_{010},\zeta_{001},\kappa\right]=\left[\SI{405}{\kilo\pascal},\SI{2.50}{\kilo\pascal},\SI{102}{\kilo\pascal},\SI{3.915}{\mega\pascal}\right]$.}
\label{tab:sharpness}
\end{table}

\begin{figure}[t]
\centering
\includegraphics[width=170mm]{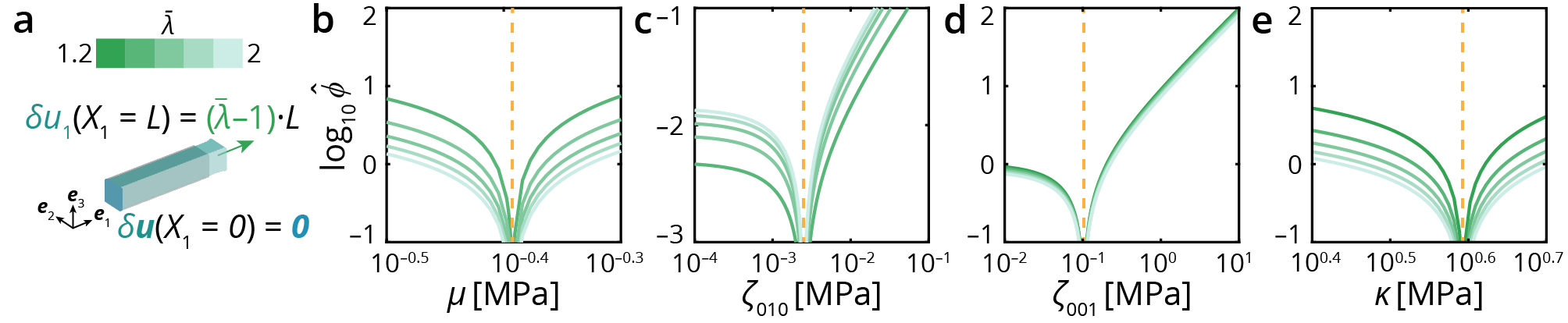}
\caption{\textbf{Comparison of cost function convergence for scaled macroscale stretch.} (a) Schematic of sample geomtery and imposed boundary displacement for increasing stretch values. (b--e) Normalized cost function profiles versus each material parameter---(b) $\mu$, (c) $\zeta_{010}$, (d) $\zeta_{001}$, (e) $\kappa$---as a function of prescribed stretch magnitude, highlight changes in parameter identifiability.}
\label{fig:phivsstretch}
\end{figure}

\medskip
\noindent
When running 3D, full-field experiments, a persistent challenge is the amount of time each individual, high-resolution volumetric scan requires.
Magnetic resonance experiments for example, as in our prior work \cite{estrada_MRu_2020, scheven_physicsMRu_2020, luetkemeyer_bioMRu_2021}, take on the order of tens of minutes at full resolution.
Hence, to compare virtual experimental sets with each other, we elected to keep the number of data sets consistent across each virtual experiment.
\Cref{fig:phivsstretch} shows how the sharpness of the estimate of each parameter is changed with increases in the maximum average stretch of the fixed-face rectangular-prismatic sample when using vSBVFs (green).
The performance for the dSBVF (blue) and aSBVF (red) techniques are shown in appendix \cref{sec:app:maxampVFs}.
In each virtual experiment, there are two intermediate steps before the final, maximum-stretch configuration.
For linear shear modulus $\mu$ (\cref{fig:phivsstretch}b) and bulk modulus $\kappa$ (\cref{fig:phivsstretch}e), an increase in the total stretch actually decreases the peak sharpness.
In contrast, $\zeta_{010}$---the coefficient associated with the $K_2^4$ term in the strain energy function (alternately, third-order in stress)---is estimated with better specificity at higher stretch values (\cref{fig:phivsstretch}c), as expected, whereas the mode-sensitive modulus $\zeta_{001}$ is relatively insensitive to the maximum average stretch magnitude (\cref{fig:phivsstretch}d).

\begin{figure}[t]
\centering
\includegraphics[width=170mm]{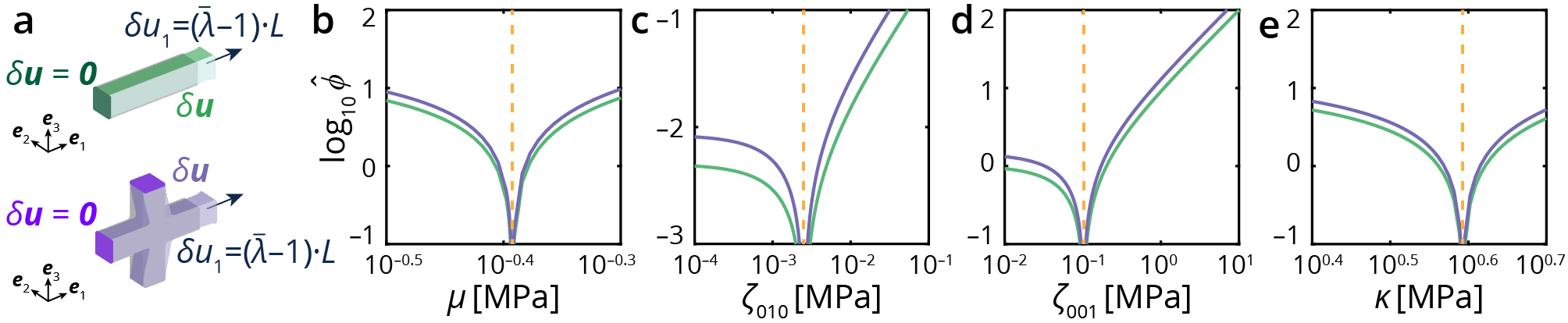}
\caption{\textbf{Cost function convergence for quasi-single-mode vs. multi-modal kinematic deformations.} (a) Geometric and boundary condtions for a quasi-uniaxial (green) test. (b--e) Cost function profiles for (b) $\mu$, (c) $\zeta_{010}$, (d) $\zeta_{001}$, and (e) $\kappa$ with quasi-single-mode deformations. (f) Schematic for a multi-model (purple) cruciform sample with both vertical and horizontal fixed surfaces. (g--j) Cost function profiles for the same parameters under multi-modal loading.}
\label{fig:phivsBVP}
\end{figure}

\begin{figure}[t]
\centering
\includegraphics[width=85mm]{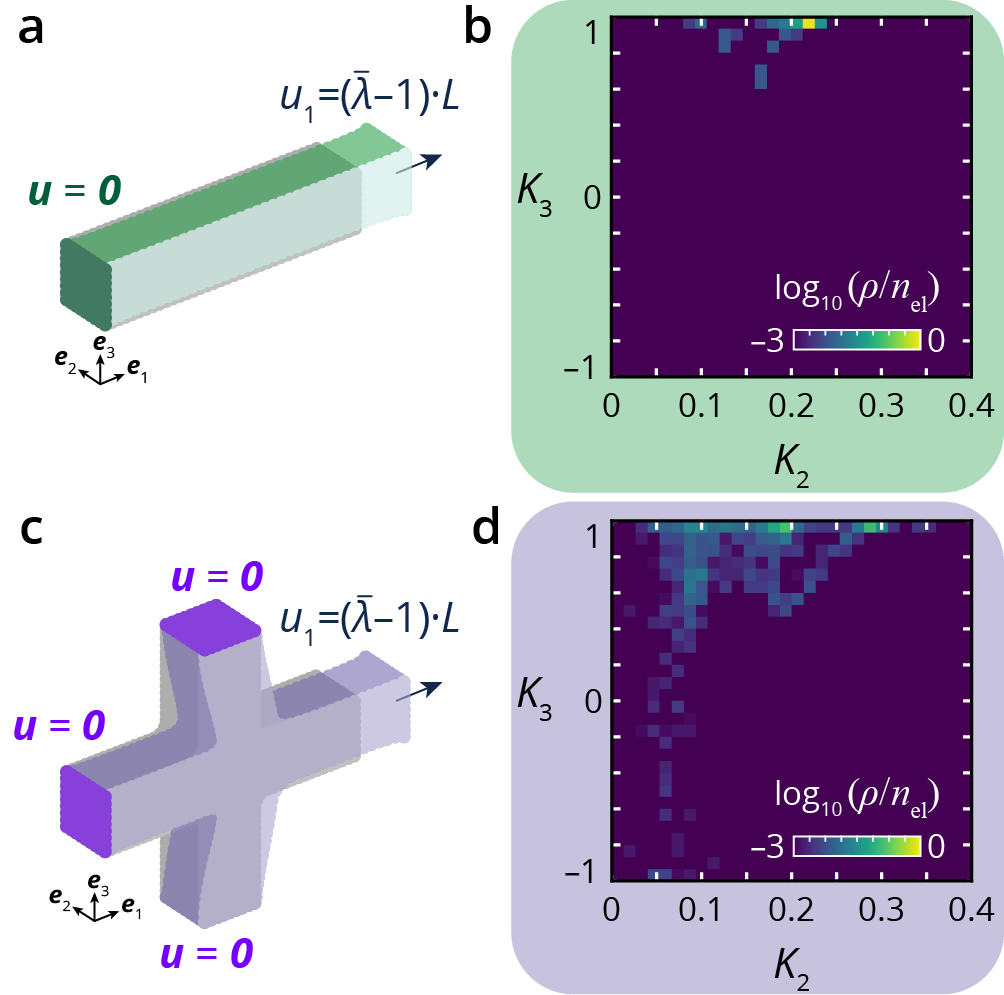}
\caption{\textbf{Kinematic diversity in full-field data for quasi-single-mode (green) and multi-model (purple) deformations} (a) Boundary conditions for a uniaxial rectangular boundary value problem. (b) 2D histogram of kinematic states in $K_2-K_3$ space for the rectangular sample. (c) Modified biaxial cruciform boundary condition. (d) 2D histogram illustrating broader spread of local kinematic states in the cruciform sample.}
\label{fig:ndhists}
\end{figure}

\medskip
\noindent
Selectivity for material parameters was compared between two sample geometries that were each actuated with a uniaxial prescribed deformation as shown in \cref{fig:phivsBVP}(a).
The geometric difference between the stretched rectangular and plus-shaped samples manifests in the different representation of kinematic states within the volumes as shown in \cref{fig:ndhists}.
The stretched rectangular sample predominantly exhibits local states of uniaxial tension (i.e. $K_3\approx 1$), whereas the actuated plus-shaped sample contains regions of uniaxial tension and/or shear.
The selectivity of material properties becomes generally better for the sample with a more varied representation in kinematic space (\cref{fig:phivsBVP}(b-e)), with sharpness values increasing by approximately 50\% for every parameter (\cref{tab:sharpness}).

\medskip
\noindent
We now consider the effects of significant experimental noise on the performance of each type of SBVFs.
Gaussian noise is added independently to the nine components of the local deformation gradient tensor $\up{F}(\bX)$ as a fraction of the maximum $u_1(\bX)/L$,
\begin{equation}
\tilde{F}_{ij}(\bX) = F_{ij}(\bX) + a \frac{\max (u_1(\bX))}{L} N(\bX),
\end{equation}
where $a$ is a constant user-chosen noise amplitude and $N(\bX)$ is randomly selected from a standard normal distribution.
Random number generator seeds were generated randomly and stored to permit comparison at the tested amplitude levels of $[0.01, 0.02, 0.03, 0.04]$.
The geometry for this comparison is a modified biaxial cruciform specimen taken to be two 40$\times$8$\times$\SI{8}{\mm} rectangular shapes overlapping and perpendicular to each other with a virtual experimental actuation of \SI{8}{\mm} along one end while the other three ends are fixed.

\begin{figure}[p]
\centering
\includegraphics[width=85mm]{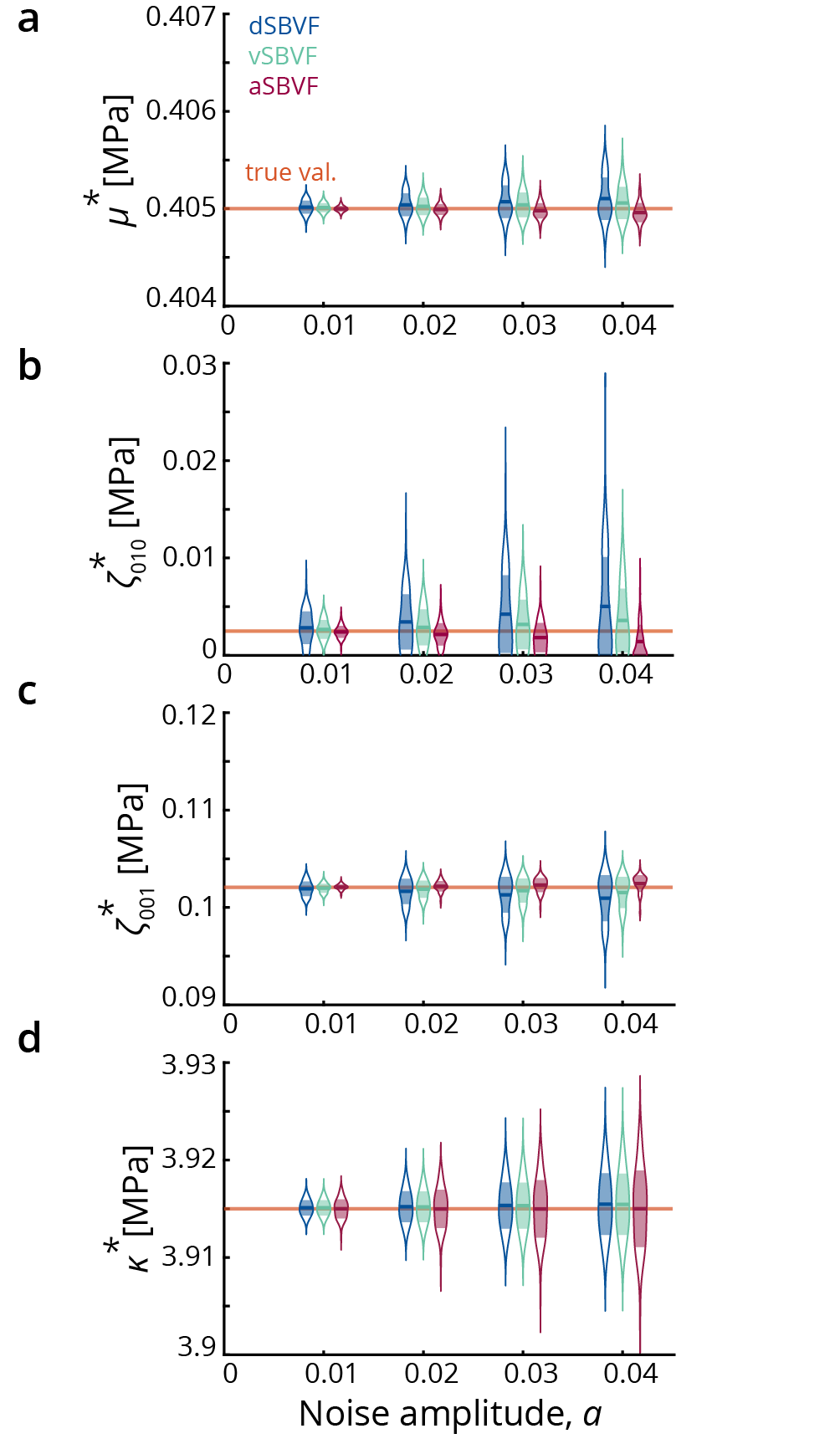}
\caption{\textbf{Sensitivity-based virtual field method comparison under additive Gaussian noise.} (a--d) Violin plots showing distributions of best-fit material parameters from 100 noisy trials for SBVF techniques---(a) shear modulus $\mu$, (b) non-linear stiffening modulus $\zeta_{010}$, (c) mode-dependent modulus $\zeta_{001}$, (d) bulk modulus $\kappa$. Means (--) and standard deviations (shaded) are shown, along with the true material property values (orange).}
\label{fig:violin}
\end{figure}

\medskip
\noindent
For all tested noise amplitudes, SBVFs universally improved both the accuracy and precision of VFM best-fit material properties in comparison to UBVFs.
We thus highlight the noisy-data performance of each of the three types of SBVF---discrete (blue), variational (sea green), and analytical (red)---for the natural strain invariant model parameters in \cref{fig:violin}.
For each amplitude, $n=100$ seeded instances of noise were added to the deformation gradient tensor $\up{F}(\bX)$.
The histograms (violins) of the best-fit ($*$) solutions illustrate the approximate linear increase of standard deviation $\sigma$ (shaded regions) with noise amplitude, even for the case of a parameter near a zero-bound.
Standard deviation values were fit to a direct linear function of noise amplitude; these values are shown in \cref{tab:std_dev}.
For the tested geometry, while the mean parameter estimates of variational SBVFs were generally (albeit non-significantly) more accurate with respect to the true values (orange line), the standard deviations of the best-fits were notably better than those for dSBVFs across all parameters.

\begin{table}[t]
\centering
\begin{tabular}{l|l l l|l l l|l l l l}
\toprule
& \multicolumn{3}{c}{\textbf{M-R}} & \multicolumn{3}{c}{\textbf{M-R (mixture)}} & \multicolumn{4}{c}{\textbf{Isotropic natural strain}} \\
\midrule
Parameter & $C_{10}$ & $C_{01}$ & $\kappa$ & $\mu$ & $\alpha$ & $\kappa$ & $\mu$ & $\zeta_{010}$ & $\zeta_{001}$ & $\kappa$ \\
\midrule
VFs & \multicolumn{10}{c}{slope of $\sigma_{\xi_i}(a)/\xi_i^*$} \\
\midrule
dSBVF & 1.667 & 46.02 & 0.2102 & 0.1028 & 45.92 & 0.2102 & 0.1377 & 528.8 & 5.971 & 0.2022 \\
vSBVF & 0.8307 & 23.34 & 0.2129 & 0.07123 & 23.27 & 0.2129 & 0.1041 & 339.1 & 4.036 & 0.2010 \\
aSBVF & -- & -- & -- & -- & -- & -- & 0.06380 & 190.8 & 2.270 & 0.2506 \\
\bottomrule
\end{tabular}
\caption{Linear relationship between normalized standard deviation of converged material properties and noise scalar quantity.}
\label{tab:std_dev}
\end{table}

\begin{table}[t]
\centering
\begin{tabular}{L|L L L L}
\toprule
\text{VFs} & \varsigma_\mu & \varsigma_{\zeta_{010}} & \varsigma_{\zeta_{001}} & \varsigma_\kappa \\
\midrule
\text{dSBVF} & 2.199\pm1.637 & 0.03242\pm0.09677 & 0.02171\pm0.06103 & 0.03252\pm0.03917 \\
\text{vSBVF} & 2.320\pm1.647 & 0.02791\pm0.04954 & 0.04034\pm0.06164 & 0.04999\pm0.03509 \\
\text{aSBVF} & 2.377\pm1.216 & 0.05571\pm0.05813 & 0.04103\pm0.02985 & 0.05336\pm0.01887 \\
\bottomrule
\end{tabular}
\caption{Cost function sharpness mean and standard deviation for 100 noise trials at a scalar noise amplitude of $a = 0.04$. In all cases, properties in forward simulations were set to $\bm{\xi}=\left[\mu,\zeta_{010},\zeta_{001},\kappa\right]=\left[\SI{405}{\kilo\pascal},\SI{2.50}{\kilo\pascal},\SI{102}{\kilo\pascal},\SI{3.915}{\mega\pascal}\right]$.}
\label{tab:sharpness_noise}
\end{table}

\section{Discussion}

When evaluating simulated, noise-free full-field data, the VFM is expected to converge to the prescribed material parameters $\bm{\xi}$.
Given the agreement of each method, i.e., $\bm{\xi}^* = \bm{\xi}$, we thus use a metric to compare the convergence of $\hat{\phi}$ around the minimum.
Specifically, we use the maximum of the Jacobian derivative $\varsigma_{\xi_i}$ (\cref{eq:sharpness}); a higher value indicates steeper, sharper convergence around $\bm{\xi}^*$.
The sharper green curves in \cref{fig:gen_v_sbvf} and higher SBVF values in \cref{tab:sharpness} illustrate the benefit of using SBVFs over UDVFs for nearly all material parameters.
As SBVFs are tailored to each material parameter, we expect and observe convergence benefits for parameters governing non-linear behavior.
While the original SBVF approach \cite{marek_sensitivityVFM_2017} is implemented in commercially available software such as MatchID (Leuven, Belgium), extensions such as vSBVFs and aSBVFs are likely to further improve higher-order parameter identification.

\medskip
\noindent Parameter identifiability can be improved by either increasing macroscale stretch to hyperelastic ranges or increasing the complexity of the BVP.
Larger macroscale deformations are useful for characterizing non-linear material parameters, while lower stretch tests are important for calibrating initial moduli (i.e., $\mu_0$ and $\kappa_0$).
As shown in \cref{fig:phivsstretch}, increasing the prescribed stretch per timestep sharpens identification of $\zeta_{010}$, but decreases the sensitivity for $\mu$ and $\kappa$.
Therefore, we expect the optimal identification of all parameters to require a dataset spanning both small and large deformations, either by sampling a range of macroscale stretch values or by designing samples that experience a spatially varying range of deformation states.
When data acquisition time is limited, as in MR-based experiments, it may not be possible to obtain many timesteps.
Our method does not require closely spaced timesteps; it analytically determines virtual fields for each parameter at any deformation state.
The overall result is efficient calibration, even in the case of restricted temporal resolution.

\medskip
\noindent
When sample design is possible, increasing the BVP complexity can further reduce the need for many timesteps.
As shown in \cref{fig:phivsBVP}, increasing the kinematic complexity of a BVP improves indentifiability across all material parameters, but only for SBVFs.
Thus, user design of kinematically rich specimens via simple modifications \cite{kim_pierronShapeOpt_2014,jones_DshapedVFM_2018,stainier_biaxialholes_2019,pierron_mt2_2021} or topological optimization \cite{barroqueiro_topopt_2020,ihuaenyi_Mechanicsinformatics_2024} stands to improve calibration, but will do so most effectively when paired with SBVFs.

\medskip
\noindent As for implementing the mechanical calibration via the SBVFM, the analytical vSBVFs and aSBVFs outperform dSBVFs for parameter identification using full-field data in the presence of noise.
\Cref{fig:violin} shows a decrease in variance of converged shear moduli $\mu$, $\zeta_{010}$ and $\zeta_{001}$ across one hundred instances with synthetically added random noise for both analytical SBVF methods (vSBVFs, and aSBVFs), compared to the discrete approach.
The variability of $\kappa$ remains consistent across all SBVF techniques.
To further quantify the differences between each method, \cref{tab:std_dev} shows the linear relationship between noise scale and variance.
Additional investigation at an amplitude level of 0.08 show that the linear behavior continues at high levels of noise.
In the dSBVF approach, stress variation $\delta\tilde{\bm{\Uppi}}$ increases variance because the difference between two timesteps of noisy data affect each virtual field evaluation, making accurate convergence more challenging.
The vSBVF and aSBVF approaches don't use timestep differences and use full-field data from one timestep to construct virtual fields.
Full-field displacement-encoding techniques, such as displacement-encoded magnetic resonance imaging \cite{aletras_DENSE_1999,neu_dense_2008} or digital volume correlation techniques with computed tomography \cite{roux_DVCCT_2008}, and X-ray scans \cite{bay_dvcorig_1999}.
By using SBVFs with directly computed derivatives, the number of time steps required for full-field data acquisition is minimized.
The implementation of the derivative-based approach is limited to identifying parameters of orthogonally decoupled models, although the variational SBVFM extends to all material models.

\section{Conclusion}
We present a significant advancement in the calibration of hyperelastic material models through full-field, three-dimensional displacement data by introducing variation-based and analytical derivative-based SBVFs.
While SBVFs substantially improve both the convergence and robustness of parameter estimation compared to traditional or user-defined virtual field strategies, 
this improvement is most pronounced for highly nonlinear model parameters and in scenarios involving measurement noise by using the derivative methods presented herein. 

\medskip
\noindent A principal outcome of this work is the systematic demonstration that all SBVF strategies reliably achieve parameter convergence under ideal (noise-free) conditions.
Moreover, vSBVFs and aSBVFs maintain sharpness and resist error even in noisy environments. 
Quantitative analysis reveals that increasing the richness of kinematic data, by optimizing sample design and enhancing macroscale stretch, consistently enhances parameter identifiability.
These findings are particularly relevant for parameters linked to higher-order material nonlinearities, which are most susceptible to error in conventional approaches.

\medskip
\noindent Our approach offers experimentalists and practitioners a flexible framework for robust, automated, and noise-tolerant material characterization. 
The vSBVF and aSBVF methodologies not only reduce reliance on manual virtual field selection but also lay the groundwork for automating future experimental mechanics workflows. 
This will be especially valuable in settings with experimental constraints, such as limited spatial and temporal resolution.

\medskip
\noindent Some limitations remain. 
While this study focused on analytically constructed SBVFs and noise characteristics typical of modern volumetric imaging modalities, further development is needed to extend these methods to more complex material behaviors, including anisotropy and viscoelasticity. 
Nonetheless, by rigorously quantifying the improvements enabled by vSBVFs and aSBVFs and providing guidance for maximizing kinematic data richness, this work marks a substantial step toward accurate, efficient, and generalizable material parameter identification in experimental mechanics.

\appendix

\section{Solution of vSBVF for the Mooney--Rivlin
and isotropic natural strain models
}
\label{sec:app:variation_approach_solution}

\noindent
In this section, we first review the derivation of the stiffness $\stiffA$ for the vSBVFs. We then discuss the reliance of this procedure on suitable choices of stress measure and volumetric strain energy function.
\newline

\noindent 
We consider the Mooney--Rivlin material model defined through the strain energy functions
\begin{align}
    \psi_{\rm iso} &= c_1 \left(\barI_1 - 3\right) + c_2 \left(\barI_2 - 3\right) \, , &
    \psi_{\rm vol} &= \frac{\kappa}{2}\left( \ln\left(J\right)\right)^2 \, ,
\label{eq:mr_energy}
\end{align} 
which results in the rotated Kirchhoff stress
\begin{equation}
    \rotkir = \kappa \ln\left(J\right) \up{I} + 
    2 c_1 \left(\overline{\up{C}} - \frac{\barI_1}{3}\up{I}\right) + 2 c_2 \left(\barI_1 \overline{\up{C}} - \overline{\up{C}}^2 - \frac{2 \barI_2}{3} \up{I} \right) \, .
\end{equation}
Taking the first variation of $\rotkir$ with respect to a virtual strain $\varu$, we have 
\begin{equation}
\begin{aligned}
    \delta_{\bu} \rotkir 
    =& \kappa  \left(\grad_{\bx}\cdot\varu\right) \up{I} 
    + \frac{2 c_1}{J^{2/3}}  \left( \delta{\up{C}} - \frac{1}{3} {\trace{\delta\up{C}} \up{I}} - \frac{2}{3} {\left(\grad_{\bx}\cdot\varu\right)\left(\up{C} - \frac{I_1}{3}\up{I}\right)}  \right) \\
    &+ \frac{2 c_2}{J^{4/3}} \Bigg{(} 
     \trace{\delta\up{C}}\up{C} + I_1 \delta\up{C} - \delta\left( \up{C}^2 \right) 
     - \frac{2}{3} \left( I_1 \, \trace{\delta\up{C}} - 2 \, \trace{\up{CF}^{\transpose} \delta{\up{F}}} \right) \up{I} \\
    & \qquad - \frac{4}{3} \left(  \grad_{\bx}\cdot\varu\right)\left(I_1 \up{C} - \up{C}^2 - \frac{2}{3} I_2 \, \up{I} 
   \right)
  \Bigg{)} \, .
\end{aligned}
\end{equation}
To satisfy $\delta_{\bu}\rotkir = \stiffA \, \delta{\up{E}_{\rm G}}$, we obtain $\stiffA = \stiffA_{\rm MR}$, where
\begin{equation}
\begin{aligned}
    \stiffA_{\rm MR}
    = & 
    \kappa \left(
      \up{I} \otimes \up{C}^{-1}
    \right)
    + \frac{4 c_1}{J^{2/3}} \left( \IIsym - \frac{1}{3}\left(\up{I}\otimes\up{I}\right) - \frac{1}{3} \left( \up{C} - \frac{I_1}{3} \up{I} \right) \otimes \up{C}^{-1} \right) \\
    & + \frac{4 c_2}{J^{4/3}} \left( \up{C}\otimes\up{I} + I_1 \IIsym - \left( \up{C} \odot \up{I} + \up{I} \odot \up{C} \right) + \frac{2}{3} \up{I} \otimes \left( \up{C} - I_1 \up{I} \right) 
    - \frac{2}{3} \left( I_1 \up{C} - \up{C}^2 - \frac{2 I_2}{3} \up{I} \right) \otimes \up{C}^{-1}
    \right)
    \, .
\end{aligned}
\end{equation}
Here, we define the operation $\odot$ between two second-order tensors $\up{Y}$ and $\up{Z}$ to result in a fourth-order tensor with components
\begin{equation}
    \left( \up{Y} \odot \up{Z} \right)_{ijkl} = \frac{1}{2}\left( Y_{ik}Z_{jl} + Y_{il}Z_{jk} \right).
\end{equation}
The fourth-order identity tensor over the space of symmetric tensors, defined as $\IIsym = \up{I} \odot \up{I}$, maps any symmetric second-order tensor to itself.
Following the procedure outlined in \ref{sec:SBVFvarbased}, we then solve for $\varu$ satisfying \cref{eq:matchVariations}, such that the variation of $\rotkir$ with respect to the displacement field is equal to that with respect to a material parameter of interest.
\medskip

\noindent
For the isotropic natural strain model defined through the strain energy functions \cref{eqn:crisland,eq:vol_strain_energy}, we similarly seek a fourth-order stiffness tensor satisfying
\begin{align}
\label{eq:matchVariations_crisland}
    \delta_{\xi_j} \rotkir &= \delta_{\bu} \rotkir = \stiffA \, \delta \mathencky
\end{align}
where $\mathencky = {\up{R}}^{\transpose} \spathencky {\up{R}}$ is the material Hencky strain. 
Notably, $\rotkir$ is the stress conjugate to $\mathencky$~\cite{hoger_henckyConjugate_1987}. 
\medskip

\noindent
Taking advantage of the simplifying assumption in \cref{eq:no_virtual_rotation_assumption}, we obtain
\begin{align}
\begin{aligned}
     \delta_{\bu}\rotkir =& \left(\kappa \delta K_1 - 2\sqrt{6}\,\zeta_{001} K_2 \delta{K_2}\right) \up{I} + \left(2\mu + 4 \, \zeta_{010} K_2^2 \right) \left(\delta\devH\right) \\
     &+ \left(8 \, \zeta_{010} K_2 \delta K_2 \right)\devH + 6\sqrt{6} \, \zeta_{001} \left(\delta \devH \right) \devH \, ,
\end{aligned}
\end{align}
where 
\begin{align}
    \devH &= {\rm dev}\left[ \mathencky \right] = \mathencky - \frac{1}{3} \trace{\mathencky} \up{I}
\end{align}
is the deviatoric part of the material Hencky strain. 
The corresponding fourth-order stiffness tensor for the isotropic natural strain model is found to be $\stiffA = \stiffA_{\rm NS}$, where
\begin{align}
\begin{aligned}
    \stiffA_{\rm NS} =& \kappa\left(\up{I} \otimes \up{I} \right)  + \left(2\mu + 4 \, \zeta_{010} K_2^2 \right) \left(\IIsym - \frac{1}{3}\up{I}\otimes\up{I}\right) \\
   &+ \left(8 \, \zeta_{010} \right)\left( \devH\otimes\devH\right) + 6\sqrt{6} \, \zeta_{001} \left( \devH\odot\up{I} - \frac{1}{3} \devH \otimes \up{I} - \frac{1}{3} \up{I} \otimes \devH \right) \, .
\end{aligned}    
\end{align}
\newline

\noindent
We note that the choice to develop vSBVF with $\rotkir$ as a stress measure is not arbitrary. Although an alternative form of $\stiffA$ can be derived according to the variation of another stress measure, the resulting $\stiffA$ would not be guaranteed to be well-conditioned for all deformation states. As an illustrative example, consider the second Piola--Kirchhoff stress, $\pktwo$, in a Mooney--Rivlin material defined by \cref{eq:mr_energy},
\begin{align}
    \pktwo &= J \, \up{F}^{-1}\cauchy\up{F}^{-\transpose} 
    = J^{-2/3} \left( \kappa \ln\left(J\right) \overline{\up{C}}^{-1} + 
    2 c_1 \left( {\up{I}} - \frac{\barI_1}{3} \overline{\up{C}}^{-1} \right) + 2 c_2 \left(\barI_1 {\up{I}} - \overline{\up{C}} - \frac{2 \barI_2}{3} \overline{\up{C}}^{-1} \right)
    \right) \, .
\end{align}
To satisfy $\delta_{\bu}\pktwo = \stiffA \, \delta{\up{E}_{\rm G}}$, we may follow the above steps and obtain $\stiffA = \stiffA_{\rm MR}^{\pktwo}$, where
\begin{equation}
\begin{aligned}
    \stiffA_{\rm MR}^{\pktwo}
    = & 
    \kappa \left( \left( \up{C}^{-1} \otimes \up{C}^{-1} \right) - 2 \ln(J) \left(\up{C}^{-1} \odot \up{C}^{-1}\right) \right) \\
    -& \frac{4 c_1}{3 J^{2/3}} \left(
    \left(\up{I} - \frac{I_1}{3}\up{C}^{-1} \right)\otimes \up{C}^{-1} 
    + \left(\up{C}^{-1} \otimes \up{I} \right) 
    - I_1 \left( \up{C}^{-1} \odot \up{C}^{-1} \right) 
    \right) 
    + \frac{4 c_2}{J^{4/3}} \left({ \left(\up{I}\otimes\up{I}\right) - \IIsym }\right) \\
    -& \frac{8 c_2}{3 J^{4/3}} \left( 
    \left(I_1 \up{I} - \up{C} - \frac{2 I_2}{3} \up{C}^{-1} \right) \otimes \up{C}^{-1} 
    + I_1 \left(\up{C}^{-1} \otimes \up{I} \right)
    - \left( \up{C}^{-1} \otimes \up{C} \right)
    - I_2 \left( \up{C}^{-1} \odot \up{C}^{-1} \right)
    \right)
    \, .
\end{aligned}
\end{equation}
When $\up{F} = J^{1/3} \, \up{I}$, corresponding to a purely volumetric deformation state,
\begin{equation}
\begin{aligned}
    \stiffA_{\rm MR}^{\pktwo}
    = & 
    J^{-4/3} \left( 
    \kappa  \left( \left( \up{I} \otimes \up{I} \right) - 2 \ln(J) \, \IIsym \right) 
    + 4 \left( c_1 + c_2 \right) \left( \IIsym - \frac{1}{3} \up{I} \otimes \up{I} \right) 
    \right)
    \, .
\end{aligned}
\label{eq:mr_AA_pk2}
\end{equation}
When $J = \exp(3/2) \approx 4.5$, $\stiffA$ becomes a scalar multiple of the projection tensor $\left( \IIsym - (1/3) \, \up{I} \otimes \up{I} \right)$. The project tensor, which maps a second-order tensor to its deviatoric part, is not invertible. In addition to the purely volumetric deformation case described above, the general form of $\stiffA_{\rm MR}^{\pktwo}$ shown in \cref{eq:mr_AA_pk2} becomes ill-conditioned at other realistic states of deformation. In these cases, we cannot accurately recover $\delta\up{E}_{\rm G}$ from \cref{eq:vSBVF_inversion}.
\newline

\noindent
We intentionally elected to work with a stress measure defined in the material description. 
To obtain the first variation of a function in the spatial description, such as $\up{B}$ and $\spathencky$, additional pull-back and push-forward operations are necessary, which would further complicate the steps needed to obtain $\stiffA$~\cite{holzapfel_nonlin_mech_2000}.
\newline

\noindent
A suitable form of $\psi_{\rm vol}$ is also critical to the performance of the vSBVF. 
For example, consider a Mooney--Rivlin material with
\begin{align}
    \psi_{\rm vol} &= \frac{\kappa}{2}\left( J - 1 \right)^2
\end{align}
instead of the form presented in \cref{eq:mr_energy}. 
In our attempt to satisfy $\delta_{\bu}\rotkir = \stiffA \, \delta{\up{E}_{\rm G}}$ according to this form of strain energy function, we would arrive at a form of $\stiffA$ that is not invertible for a purely volumetric deformation with $J = 1/2$.

\section{Cost function sharpness with respect to macroscale deformation across all choices of SBVF techniques}
\label{sec:app:maxampVFs}

\begin{figure}[t]
    \centering
    \includegraphics[width=170mm]{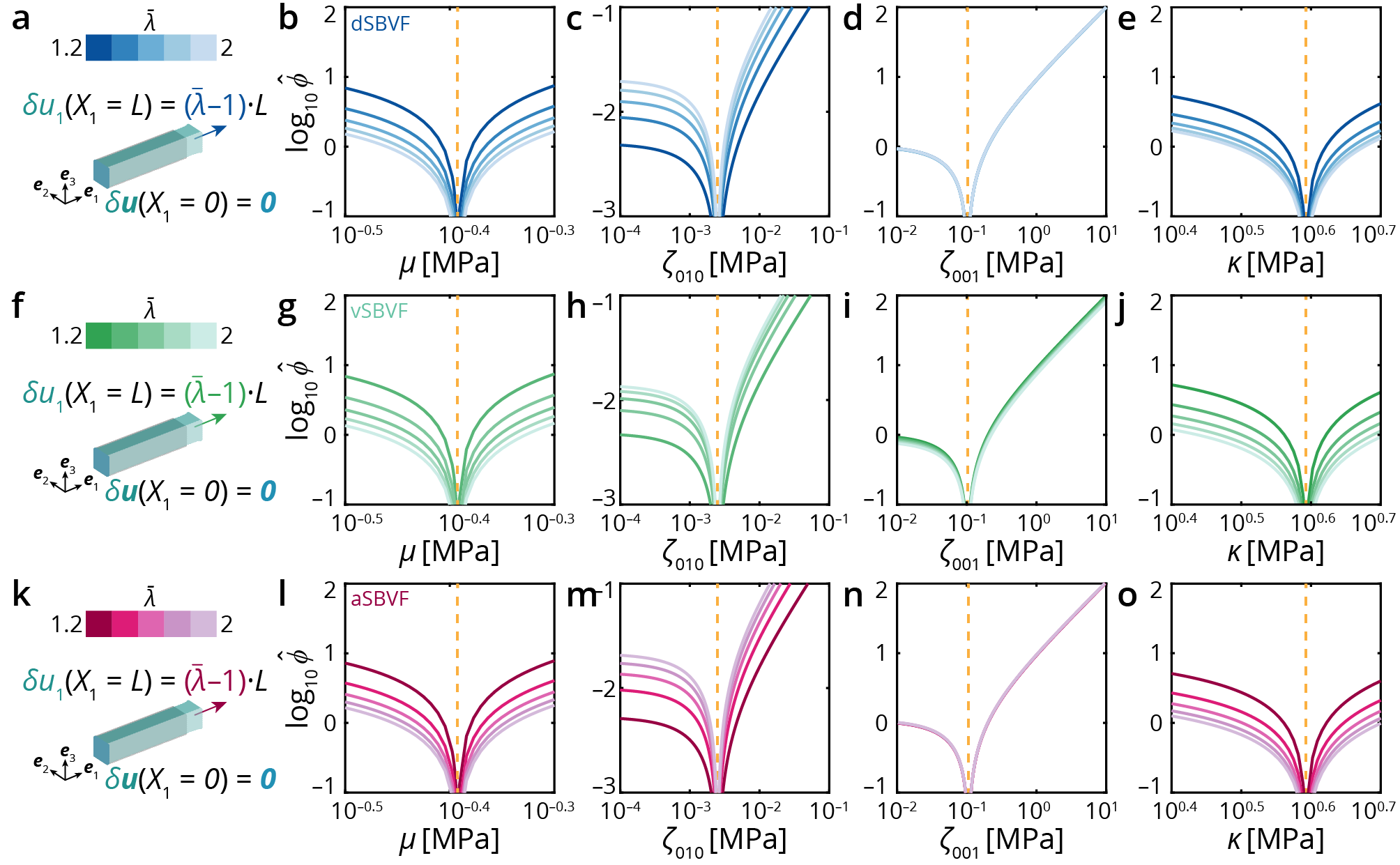}
    \caption{\textbf{Cost function sharpness as a function of macroscale deformation for different SBVF methods.} (a,f,k) Sample geometris and boundary conditions for dSBVF, vSBVF, and aSBVF tests, respectively. (b--e) Cost function sharpness for dSBVFs across parameters: (b) shear modulus $\mu$, (c) non-linear stiffening modulus $\zeta_{010}$, (d) mode-dependent modulus $\zeta_{001}$, and (e) bulk modulus $\kappa$. (g--j) Results for vSBVF methods. (l--o) Results for aSBVF methods. Analyses are shown for increasing prescribed displacement.}
    \label{fig:phivsstretch_app}
\end{figure}

\section*{Data Access}
All data will be made available upon acceptance with a DOI to the Deep Blue Data repository hosted at the University of Michigan. 
All data in the repository are available for anyone to download without restriction and are licensed under a Creative Commons 4.0 license.

\section*{Funding}

\textbf{DPN} acknowledges a Rackham Graduate Fellowship from the University of Michigan. 
\textbf{JBE} acknowledges NSF CAREER award CMMI2338371.

\section*{Author Contributions}

\textbf{DPN}: conceptualization, methodology, software, validation, formal analysis, investigation, writing – original draft, writing – review and editing, visualization. 
\textbf{ZZ}: conceptualization, methodology, software, validation, formal analysis, investigation, writing - original draft, writing - review and editing.
\textbf{JBE}: conceptualization, methodology, validation, writing – original draft, writing – review and editing, supervision, project administration, funding acquisition. 

\section*{Competing Interests}
The authors declare no competing financial interests or personal conflicts of interest that could have influenced the work described in this paper.

\printbibliography

\end{document}